\documentclass[journal]{IEEEtran}

\usepackage{amsmath, amssymb}
\usepackage{graphicx}
\usepackage{hyperref}
\usepackage{geometry}
\usepackage{booktabs}
\usepackage{float}

\title{AI-Native Autonomous Infrastructure (ANAI): A Formal Framework for the Next General-Purpose Technology}

\author{
Hidir Selcuk Nogay\\
Bursa Uludag University\\
Department of Electrical and Electronics Engineering\\
Bursa, Türkiye\\
\texttt{hsnogay@uludag.edu.tr}
}

\date{}

\begin{document}

\maketitle

\begin{abstract}
Artificial intelligence is increasingly described as a candidate next-generation general-purpose technology (GPT). However, existing interpretations predominantly emphasize performance scaling rather than structural transformation. This paper introduces a formal framework for evaluating AI as a systemic infrastructural transition rather than merely a computational breakthrough. We propose the concept of AI-Native Autonomous Infrastructure (ANAI), defined as a regime in which decision-autonomy becomes embedded within critical infrastructures.
The framework operationalizes this transition through three quantitative constructs: the Autonomy Index (AIx), the Infrastructure Coupling Coefficient (ICC), and the Technological Transition Potential (TTP). We formalize the joint scaling dynamics of autonomy and infrastructural embedding, derive threshold conditions for paradigm transition, and introduce a phase-space representation of systemic transformation. A temporal transition model further illustrates how nonlinear co-evolution between autonomy and infrastructure integration produces super-linear growth in transition potential.
Unlike prior GPT cycles, the ANAI regime exhibits a recursive energy–computation feedback loop in which AI systems both increase computational demand and optimize the infrastructures that sustain them. This feedback mechanism accelerates infrastructural embedding and differentiates AI-driven transformation from previous technological revolutions.
By shifting analytical focus from model performance to infrastructural autonomy and coupling intensity, this study offers a conceptual and mathematical foundation for assessing whether artificial intelligence constitutes the next general-purpose technology.

\end{abstract}
\section{Introduction}

Artificial intelligence (AI) is increasingly described as a transformative technology with the potential to reshape economic, industrial, and social systems. Much of the contemporary discourse, however, focuses primarily on performance scaling—model size, parameter growth, benchmark accuracy, and computational efficiency. While these metrics capture rapid technical progress, they do not by themselves determine whether AI constitutes a structural technological revolution comparable to historical general-purpose technologies (GPTs) ~\cite{Helpman1994GPT,Jovanovic2005GPT}.
Classical GPT theory emphasizes technologies that are pervasive, continuously improving, and capable of spawning complementary innovations across multiple sectors. Steam power, electricity, computing, and the internet each satisfied these conditions by reorganizing production structures and embedding themselves into critical infrastructures. In each case, the transformative effect did not arise solely from technical capability, but from infrastructural integration and systemic diffusion ~\cite{Helpman1994GPT,Jovanovic2005GPT,Lipsey2006GPT}.
This paper argues that artificial intelligence should be evaluated not merely as a computational breakthrough, but as a candidate infrastructural paradigm. The central question is therefore not how intelligent AI models become in isolation, but whether decision autonomy becomes structurally embedded within critical infrastructures. When autonomous decision-making systems are tightly integrated into energy grids, financial markets, logistics networks, manufacturing systems, and research pipelines, AI transitions from a tool to an infrastructural substrate ~\cite{Ganguli2022Predictability,Arora2023Emergence,Caballero2022Scaling}.
To formalize this transition, we introduce the concept of AI-Native Autonomous Infrastructure (ANAI). ANAI describes a regime in which autonomous computational systems are no longer peripheral optimizers but core coordinators of socio-technical systems. This regime differs qualitatively from prior GPT waves because autonomy itself becomes an infrastructural variable rather than an application-level feature ~\cite{Geels2004SocioTechnical,Fuenfschilling2014Structuration}.
Figure 1 situates this argument historically by contrasting previous GPT transitions with the emerging ANAI paradigm. Unlike earlier technological cycles, where energy and information infrastructures expanded separately, the AI-driven transition introduces a coupling between computation, autonomy, and infrastructure intensity ~\cite{Rubei2025EnergyLLM,Chauhan2024EnergyDemand}.
\begin{figure*}[!t]
\centering
\includegraphics[width=\textwidth]{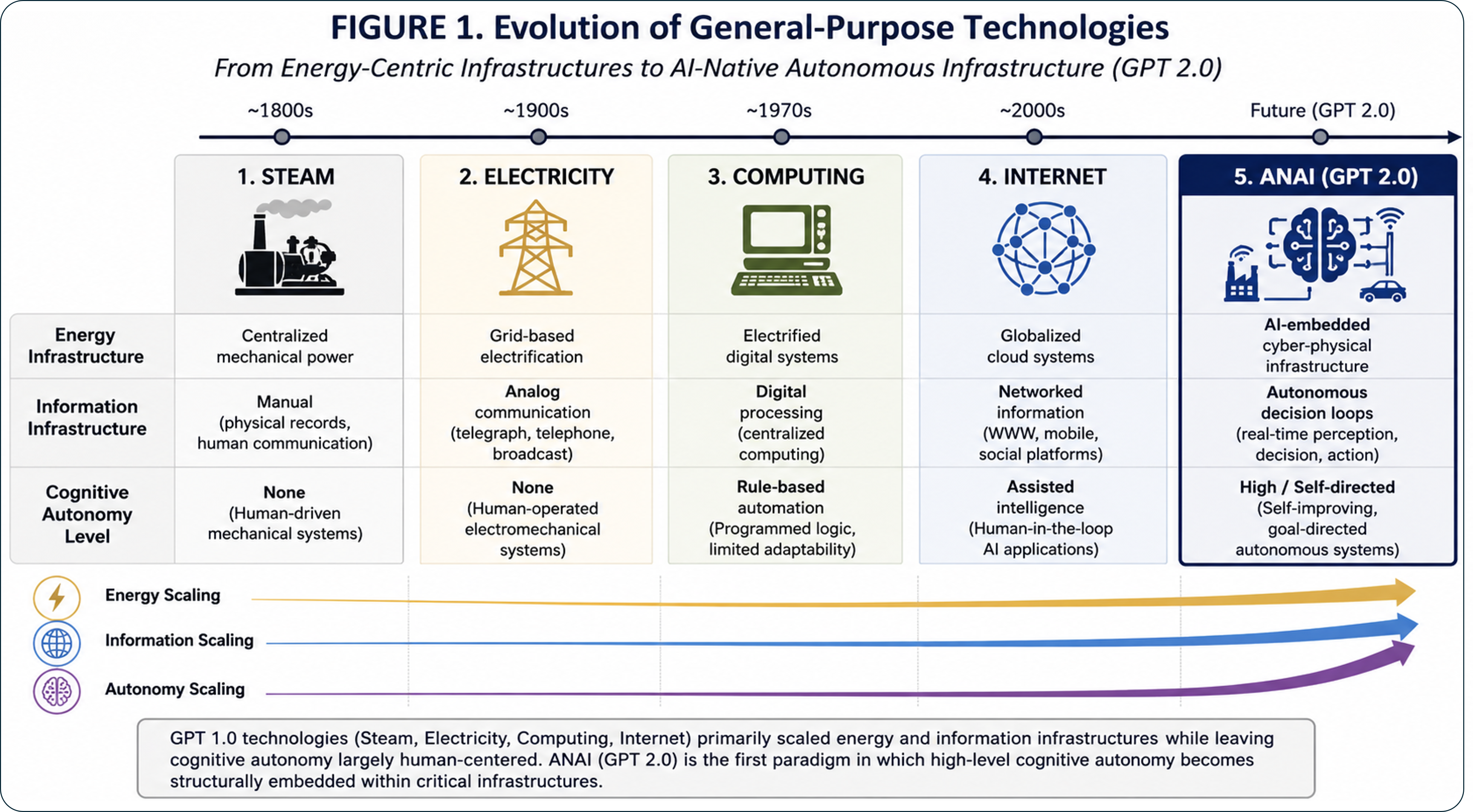}
\caption{Historical evolution of general-purpose technologies and the emergence of cognitive autonomy.}
\label{fig:figure1}
\end{figure*}
The contribution of this paper is threefold. First, we extend GPT theory to incorporate measurable decision autonomy as a structural parameter. Second, we introduce quantitative constructs—the Autonomy Index (AIx), the Infrastructure Coupling Coefficient (ICC), and the Technological Transition Potential (TTP)—to model infrastructural embedding. Third, we propose a phase-space and temporal transition framework that captures nonlinear co-evolution between autonomy and infrastructural integration.
By shifting analytical focus from model performance to infrastructural embedding, this work provides a conceptual and mathematical foundation for assessing whether artificial intelligence represents the next general-purpose technology.

\section{Formal Model of AI-Native Autonomous Infrastructure}
General-purpose technologies are widely associated with structural transformations that extend beyond direct productivity effects. Such transformations can be interpreted as paradigm shifts in the sense described by Dosi ~\cite{Dosi1982TechnologicalParadigms}, where innovation trajectories become reorganized under new problem-solving frameworks. These shifts typically entail the reconfiguration of large technical systems, in which technological components co-evolve with organizational and infrastructural structures ~\cite{Hughes1987LargeSystems}. Moreover, paradigm transitions are embedded within institutional environments that both constrain and enable systemic change ~\cite{North1990Institutions}. From this perspective, the rise of AI-native infrastructures may signal not merely technological scaling but the restructuring of an entire socio-technical configuration.
Historically, steam power, electricity, computing, and the internet have been identified as canonical GPTs ~\cite{Helpman1994GPT,Jovanovic2005GPT,Lipsey2006GPT}. In each case, the transformative impact emerged gradually through infrastructural embedding rather than immediate performance superiority. Electricity, for example, did not revolutionize manufacturing simply by replacing steam engines; its impact materialized when factory layouts were reorganized around decentralized electric motors [3]. Similarly, computing evolved from isolated calculation devices to networked information infrastructure only after widespread integration across sectors. Three structural properties consistently characterize GPT transitions: pervasive diffusion across sectors, continuous technical improvement, and complementarity-driven innovation cascades [1-3].
However, classical GPT theory treats intelligence implicitly—as an enabler of optimization—rather than as a measurable infrastructural variable ~\cite{Bresnahan2010GPT}. Artificial intelligence introduces a distinct possibility: decision autonomy itself may become structurally embedded in production and coordination systems.
Contemporary AI discussions often emphasize scaling laws, parameter growth, and benchmark performance ~\cite{Kaplan2020ScalingLaws}. While these developments indicate rapid technical advancement, GPT classification requires more than computational scale. It requires infrastructural coupling and systemic reallocation of coordination mechanisms.
In contrast to prior GPT waves, AI systems are capable of performing adaptive decision-making in real time  ~\cite{Shrestha2019OrgAI}. When deployed at scale, such systems can assume coordination roles traditionally performed by human operators. This shift represents not merely technological substitution but institutional reconfiguration.
The key theoretical gap, therefore, lies in formalizing when AI crosses the threshold from tool-like augmentation to infrastructural autonomy. Existing GPT literature provides criteria for pervasiveness and complementarity but lacks quantitative constructs for measuring autonomy intensity and infrastructural embedding. This paper addresses that gap by introducing measurable indices that extend GPT theory into the domain of autonomous computational infrastructure.
By reframing AI as a candidate infrastructural regime rather than a performance-driven innovation, we build upon classical GPT theory while proposing a formal extension capable of modeling autonomy-driven systemic transformation.

\section{Defining AI-Native Autonomous Infrastructure (ANAI)}

\subsection{Conceptual Definition}
AI-Native Autonomous Infrastructure (ANAI) refers to a socio-technical system in which artificial intelligence is not merely an application layer, but a structurally embedded decision-making core within critical infrastructures.
Unlike traditional digital infrastructures that facilitate information exchange, ANAI systems internalize algorithmic cognition into operational control loops. In such systems, AI models continuously perceive, evaluate, decide, and execute actions within economic, industrial, and social environments.
We define ANAI as a technological configuration satisfying three structural conditions:
1.	Embedded Autonomy – AI systems execute operational decisions without continuous human intervention. 
2.	Recursive Self-Optimization – Performance improves through data feedback and adaptive model updates. 
3.	Infrastructure-Level Integration – AI systems are embedded within energy grids, logistics networks, financial systems, manufacturing processes, or knowledge production systems. 
This distinguishes ANAI from earlier digital paradigms in which AI functioned primarily as decision support rather than decision execution ~\cite{Shrestha2019OrgAI,Kellogg2020Algorithms}.
Crucially, ANAI represents a shift from:
Human-in-the-loop systems → Hybrid human-AI systems → AI-centered control architectures. This progression implies that the locus of agency migrates from human operators to distributed algorithmic entities embedded in infrastructure layers.

\subsection{Formalizing Autonomy: The Autonomy Index (AIx)}
To move beyond purely conceptual characterizations of autonomy, a measurable construct is required. We define the \textbf{Autonomy Index (AIx)} as a normalized indicator that captures the degree to which operational decision authority is transferred from human agents to AI systems within a given infrastructure. Formally, the Autonomy Index is defined as:
\begin{equation}
AIx \in [0,1]
\label{eq:aix_range}
\end{equation}
where $AIx = 0$ denotes purely human-centered decision-support systems, and $AIx = 1$ represents fully autonomous, self-modifying systems operating without continuous human intervention.
To operationalize this construct, we decompose $AIx$ into four structural components:

\begin{equation}
AIx = \frac{D + E + R + M}{4}
\label{eq:aix_components}
\end{equation}
where:

\begin{itemize}
\item $D$ represents Decision Independence,
\item $E$ denotes Execution Autonomy,
\item $R$ captures Real-time Adaptive Responsiveness,
\item $M$ reflects Model Self-Modification Capacity.
\end{itemize}
Each component is normalized within the interval $[0,1]$, ensuring that the composite index remains bounded as defined in \eqref{eq:aix_range}.
Decision Independence (D) measures the proportion of operational decisions executed without human override. Execution Autonomy (E) quantifies the extent to which system actions are autonomously carried out following internal inference processes. Real-time Adaptive Responsiveness (R) captures the presence of closed-loop feedback mechanisms enabling dynamic adjustment to environmental inputs. Model Self-Modification Capacity (M) evaluates whether the system can update parameters, refine internal representations, or modify decision policies without external reprogramming.
Equation~\eqref{eq:aix_components} therefore formalizes autonomy as a multi-dimensional construct rather than a binary classification. Importantly, historical GPTs such as electricity or early computing infrastructures would yield $AIx \approx 0$, as they enhanced productivity without embedding independent decision logic. In contrast, large-scale AI-native systems increasingly exhibit non-trivial values of AIx, indicating a structural departure from previous technological paradigms.
Thus, autonomy becomes an analytically tractable dimension within GPT theory, extending classical criteria by incorporating decision agency into infrastructural analysis.

\subsection{Infrastructure Coupling Coefficient (ICC)}
Autonomy alone is insufficient to qualify a technology as a next-generation general-purpose technology. A system may exhibit high autonomy while remaining confined to a limited domain. To capture the systemic scale of technological embedding, we introduce the Infrastructure Coupling Coefficient (ICC). The ICC measures the structural integration of autonomous AI systems within critical infrastructures and is defined as:
\begin{equation}
ICC \in [0,1]
\label{eq:icc_range}
\end{equation}
Operationally, we define:

\begin{equation}
ICC = E_p \cdot D_p \cdot P_p
\label{eq:icc_components}
\end{equation}
where:

\begin{itemize}
\item $E_p$ represents Energy Infrastructure Penetration,
\item $D_p$ denotes Data Infrastructure Dependence,
\item $P_p$ captures Physical System Embedding.
\end{itemize}
The multiplicative structure in \eqref{eq:icc_components} reflects the systemic interdependence among these dimensions. If any component approaches zero, overall infrastructure coupling remains limited, preventing systemic transformation.
While Equation~\eqref{eq:icc_components} captures structural embedding, the system transition toward GPT status requires the interaction between autonomy and infrastructure coupling. We therefore define the Technological Transition Potential (TTP) as:

\begin{equation}
TTP = AIx \cdot ICC
\label{eq:ttp}
\end{equation}
A technology qualifies as a next-generation GPT when:

\begin{equation}
TTP > \tau
\label{eq:threshold}
\end{equation}
where $\tau$ represents a critical transition threshold determined by macroeconomic and institutional conditions.
Equation~\eqref{eq:ttp} highlights that systemic transformation requires both high decision autonomy and deep infrastructural embedding. High autonomy without infrastructure coupling produces localized innovation. High coupling without autonomy results in digitally enhanced but human-centered systems. Only their interaction generates the structural conditions necessary for paradigm-level technological shifts.

\section{Application of the ANAI Framework to Critical Infrastructures}
The analytical constructs introduced in Sections~3.2 and~3.3 gain explanatory power when applied to real-world infrastructures undergoing AI-driven transformation. This section illustrates how the Autonomy Index (AIx), the Infrastructure Coupling Coefficient (ICC), and the Technological Transition Potential (TTP) defined in \eqref{eq:ttp} can be interpreted across critical domains. Rather than assigning precise numerical values, we examine structural trajectories and relative positioning across sectors.

\subsection{Smart Energy Grids}
Energy systems represent one of the most structurally sensitive infrastructures in modern economies. Traditionally, grid operation relied on human-supervised control architectures, where automation supported monitoring but did not independently execute strategic decisions. Recent integration of AI-based forecasting, load balancing, and distributed generation optimization introduces measurable autonomy in grid operation~\cite{Velasquez2024SmartGrid,Acharya2025AgenticAI}. In advanced smart grid implementations:

\begin{itemize}
\item Real-time demand forecasting adjusts generation schedules autonomously.
\item Distributed renewable sources are optimized through algorithmic dispatch.
\item Storage systems execute charge-discharge cycles without manual intervention.
\end{itemize}
Within the ANAI framework, these systems exhibit moderate-to-high values of $D$, $E$, and $R$, increasing $AIx$ as defined in \eqref{eq:aix_components}. However, model self-modification capacity ($M$) remains partially constrained by regulatory oversight and safety requirements.
On the infrastructure side, energy systems score high in $E_p$ and $P_p$, since AI directly influences physical power flows. Consequently, $ICC$, as defined in \eqref{eq:icc_components}, grows significantly when AI penetrates grid management layers. The resulting $TTP$ in \eqref{eq:ttp} indicates that smart grids represent one of the leading domains approaching GPT-level transition conditions.

\subsection{Autonomous Scientific Research Platforms}
Emerging AI systems increasingly participate in literature synthesis, hypothesis generation, molecular design, and automated laboratory experimentation, embedding algorithmic cognition directly into scientific discovery pipelines~\cite{Grisoni2021GenerativeDrug,Yang2019AIDrugDiscovery,Thakkar2021AISynthesis}.
In such systems, decision independence ($D$) and adaptive responsiveness ($R$) increase as AI models iteratively refine hypotheses, experimental parameters, and evaluation criteria through closed feedback loops. Model self-modification capacity ($M$) becomes non-trivial when generative systems update internal representations based on experimental outcomes, rather than operating under fixed rule-based pipelines. Unlike traditional digital tools that merely support human-led research, these platforms embed algorithmic cognition directly into the epistemic core of scientific production, resulting in rising $AI_x$ values.
From an infrastructure perspective, coupling manifests through both data dependence ($D_p$) and physical laboratory embedding ($P_p$). As AI-driven laboratories integrate robotics, material synthesis modules, and simulation pipelines, the Infrastructure Coupling Coefficient (ICC) increases accordingly. This indicates a transition from assistive automation toward structurally embedded autonomy within epistemic infrastructures.

\subsection{Financial Clearing and Autonomous Trading Systems}
Financial systems provide a domain where algorithmic decision-making is already deeply embedded within operational infrastructures. High-frequency trading (HFT) platforms execute transactions at latencies beyond human intervention, exhibiting substantial decision independence ($D$) and execution responsiveness ($R$) \cite{Ersan2016HFT,Mandes2024HFTReview}. 

However, many such systems remain bounded by predefined strategies and regulatory rule sets rather than recursive model self-modification. As a result, while $AI_x$ may reach high levels in $D$ and $E$, its self-modification capacity ($M$) typically remains moderate. 

From an infrastructure perspective, financial AI systems exhibit extremely high data infrastructure dependence ($D_p$), as they rely on real-time market feeds, co-located servers, and low-latency execution architectures. In contrast, physical embedding ($P_p$) is comparatively lower than in energy grids or AI-driven laboratories, since material-world coupling is indirect.

This asymmetry implies that financial AI systems may achieve operational autonomy without full infrastructural entanglement. Consequently, high $AI_x$ does not necessarily imply high Infrastructure Coupling Coefficient (ICC), reinforcing the analytical necessity of jointly evaluating $AI_x$ and ICC as formalized in Equation~(5).

\subsection{Autonomous Manufacturing and Logistics Networks}

Manufacturing ecosystems increasingly integrate predictive maintenance, robotic control, supply chain optimization, and dynamic routing systems~\cite{Oztemel2020Industry40}. When AI systems coordinate procurement, production, and distribution decisions within closed-loop architectures, both $AI_x$ and Infrastructure Coupling Coefficient (ICC) rise simultaneously. Physical embedding ($P_p$) becomes significant due to robotics integration, sensor networks, and cyber-physical production systems.

Unlike financial platforms that exhibit high decision independence but moderate infrastructural embedding, manufacturing systems combine recursive decision-making, adaptive responsiveness, and physical-world execution. This dual integration positions manufacturing as a leading candidate domain for approaching the transition threshold defined in Equation~(6).

To synthesize the cross-domain analysis presented in Sections~4.1–4.3, Table~\ref{tab:table1} maps representative infrastructures within the $AI_x$–ICC phase space introduced in Figure~2. The table illustrates how different domains occupy distinct regions of the autonomy–infrastructure coupling spectrum and how their transition positioning varies accordingly.

\begin{table}[h]
\centering
\caption{Qualitative assessment of $AI_x$, ICC, and transition positioning across representative domains}
\label{tab:table1}
\begin{tabular}{lcccc}
\hline
\textbf{Domain} & \textbf{$AI_x$} & \textbf{ICC} & \textbf{TTP} & \textbf{Transition Stage} \\
\hline
Smart Grid      & 0.65 & 0.75 & 0.49 & Near ANAI \\
Finance         & 0.80 & 0.50 & 0.40 & Transitional \\
Research Labs   & 0.60 & 0.45 & 0.27 & Emerging \\
Manufacturing   & 0.80 & 0.80 & 0.64 & ANAI \\
\hline
\end{tabular}
\vspace{2mm}

\footnotesize{Values are illustrative and conceptual, intended to represent relative positioning within the ANAI framework rather than empirically measured quantities.}
\end{table}
\subsubsection*{Analytical Interpretation}

\begin{figure*}[!t]
\centering
\includegraphics[width=\textwidth]{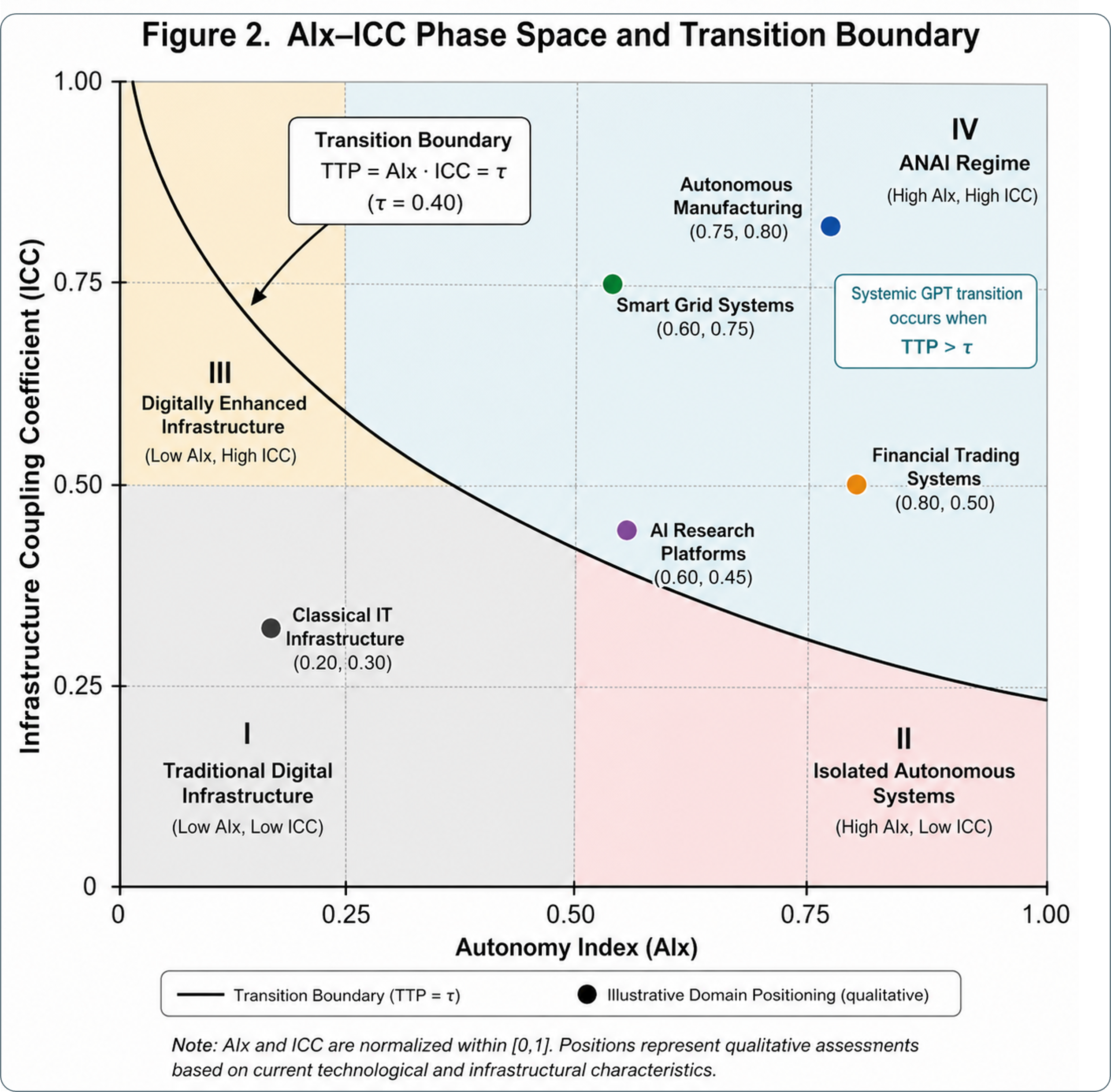}
\caption{$AI_x$–ICC phase space illustrating the ANAI transition boundary (TTP = $\tau$) and qualitative domain positioning.}
\label{fig:figure2}
\end{figure*}

As summarized in Table~\ref{tab:table1} and visually represented in Figure~\ref{fig:figure2}, manufacturing systems exhibit the strongest convergence of autonomy and infrastructure coupling, placing them closest to the ANAI regime. Across these domains, a consistent structural pattern emerges:

\begin{itemize}
    \item $AI_x$ tends to increase first in cognitively intensive domains.
    \item ICC increases when AI penetrates energy, logistics, and material infrastructures.
    \item TTP rises nonlinearly when both indices scale simultaneously.
\end{itemize}

This observation supports the central thesis that AI qualifies as a candidate next-generation GPT not merely because of performance improvements, but because of its convergence of autonomous decision-making and infrastructural embedding.

\section{Transition Dynamics of AI-Native Autonomous Infrastructure}

Technological paradigms do not emerge instantaneously. Historical GPT transitions exhibit nonlinear adoption trajectories characterized by slow incubation, rapid scaling, and eventual saturation~\cite{Bass1969GrowthModel,Rogers2003Diffusion}. To model the temporal evolution of AI-Native Autonomous Infrastructure (ANAI), we extend the static framework introduced in Section~3 into a dynamic formulation. Let autonomy and infrastructure coupling evolve over time:

\begin{equation}
AI_x = AI_x(t), \quad ICC = ICC(t)
\label{eq:dynamic_variables}
\end{equation}

We assume that both follow logistic growth dynamics, consistent with empirical models of technological diffusion.

\subsection{Logistic Growth of Autonomy}

The evolution of autonomy intensity can be modeled through a constrained logistic dynamic:

\begin{equation}
\frac{d AI_x}{dt} = \alpha \, AI_x \left(1 - \frac{AI_x}{K_A} \right)
\label{eq:logistic_autonomy}
\end{equation}

where $\alpha$ denotes the intrinsic scaling rate of autonomy and $K_A \leq 1$ represents the structural upper bound imposed by regulatory, ethical, safety, and institutional constraints. 
The logistic formulation reflects a fundamental characteristic of technological scaling. In early stages, autonomy growth remains gradual due to algorithmic immaturity, limited data availability, and restricted deployment environments. As model architectures improve and infrastructure expands, scaling accelerates, producing rapid gains in capability and operational independence. However, growth cannot continue indefinitely. Institutional governance, safety validation, and socio-political constraints introduce saturation effects, causing autonomy intensity to asymptotically approach its structural ceiling.

This bounded acceleration dynamic is essential for understanding systemic transformation. Autonomy scaling alone does not guarantee paradigm transition; rather, its interaction with infrastructural embedding determines whether localized optimization evolves into generalized technological dominance.

\subsection{Logistic Growth of Infrastructure Coupling}

Infrastructure coupling evolves according to a constrained logistic dynamic:

\begin{equation}
\frac{dICC}{dt} = \beta\, ICC \left(1 - \frac{ICC}{K_I}\right)
\end{equation}

where $\beta$ denotes the rate of infrastructural embedding and $K_I \leq 1$ represents the structural ceiling imposed by physical capacity, regulatory environments, capital allocation, and institutional constraints.

Unlike autonomy scaling, which can accelerate rapidly through software optimization and algorithmic refinement, infrastructure embedding is inherently bounded by material deployment cycles, grid stability requirements, safety validation procedures, and governance frameworks. As a result, ICC typically exhibits a delayed yet cumulative growth trajectory. The logistic formulation captures this structural inertia: early-stage expansion proceeds gradually, accelerates during coordinated investment phases, and eventually saturates as systemic constraints become binding.

This bounded growth dynamic plays a critical role in shaping the overall transition process. When infrastructural embedding lags autonomy intensity, systemic transformation remains fragmented. Conversely, synchronized scaling across both dimensions enables multiplicative acceleration in transition.

\subsection{Technological Transition Potential Over Time}

Using the multiplicative definition of Technological Transition Potential in Eq.~(5) and substituting the logistic trajectories from Eqs.~(8) and (9), a nonlinear interaction dynamic emerges. Because TTP scales multiplicatively, systemic transformation requires synchronized growth across autonomy intensity and infrastructure coupling. The paradigm transition occurs when the threshold condition defined in Eq.~(6) is satisfied.

\begin{figure*}[!t]
\centering
\includegraphics[width=\textwidth]{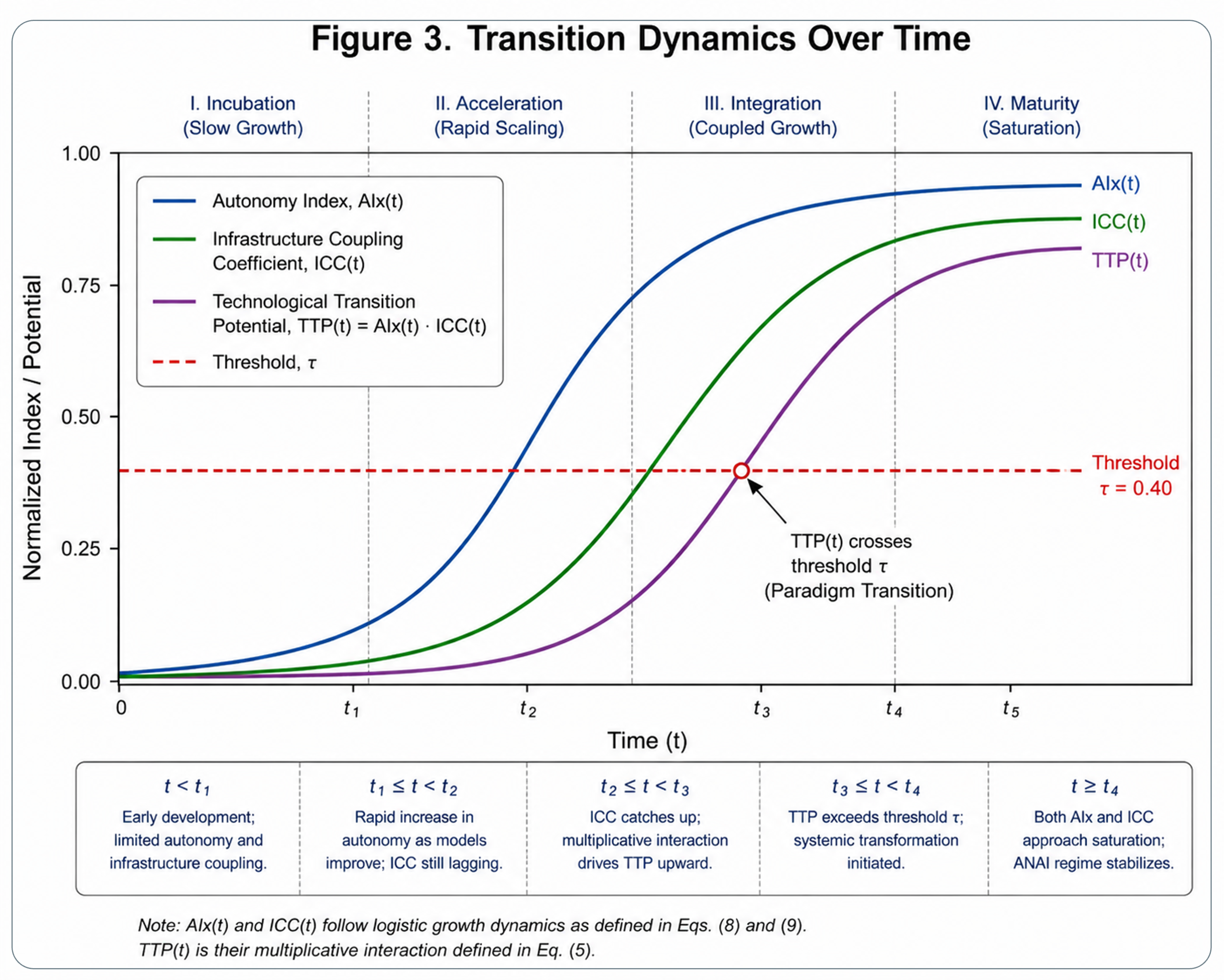}
\caption{Joint logistic evolution of autonomy and infrastructure coupling and the resulting nonlinear escalation of technological transition potential.}
\label{fig:figure3}
\end{figure*}

As illustrated in Figure~\ref{fig:figure3}, the logistic trajectories defined in (8) and (9) generate a nonlinear interaction dynamic when combined through the multiplicative structure of TTP(t) in (5). During early stages, incremental gains in autonomy or infrastructure coupling produce only modest increases in transition potential. However, once both $AI_x(t)$ and $ICC(t)$ approach intermediate levels, their interaction accelerates super-linearly, causing TTP(t) to rise sharply. The paradigm transition occurs when the trajectory crosses the threshold condition specified in (6), marking the onset of AI-Native Autonomous Infrastructure (ANAI).

This temporal interpretation highlights that systemic transformation is not driven by isolated performance improvements, but by synchronized scaling across autonomy and infrastructure embedding. Figure~\ref{fig:figure3} therefore captures the dynamic mechanism underlying GPT emergence within the ANAI framework.

\subsection{Phase-Space Interpretation}

As illustrated in Figure~\ref{fig:figure2}, the phase-space representation provides a structural interpretation of the transition condition defined in (6). The hyperbolic boundary corresponding to $AI_x \cdot ICC = \tau$ separates digitally enhanced but human-centered systems from autonomy-embedded infrastructures. Systems located below the boundary may exhibit either high autonomy or strong infrastructure integration in isolation, yet fail to produce systemic transformation. Only when both dimensions scale simultaneously does the Technological Transition Potential (TTP) defined in (5) exceed the critical threshold.

The qualitative positioning of representative domains further demonstrates that paradigm-level transition is not determined by model performance alone, but by the coordinated expansion of decision autonomy and infrastructural embedding. This phase-space perspective reframes GPT emergence as a joint scaling phenomenon rather than a unidimensional technological improvement process. As shown in Figure~\ref{fig:figure2}, high autonomy alone does not guarantee systemic transformation unless accompanied by deep infrastructural embedding.

Four regimes emerge:
\begin{enumerate}
\item Low $AI_x$ / Low ICC -- Traditional digital infrastructure
\item High $AI_x$ / Low ICC -- Isolated autonomous systems
\item Low $AI_x$ / High ICC -- Digitally optimized human-centered infrastructure
\item High $AI_x$ / High ICC -- AI-Native Autonomous Infrastructure (ANAI regime)
\end{enumerate}

The boundary separating regime (4) from the others corresponds to the threshold $\tau$ in (6). This phase-space representation suggests that the GPT transition is not driven solely by model performance gains but by coordinated growth across autonomy and infrastructure embedding.

\section{Energy--Computation Feedback and Systemic Acceleration}

Unlike previous general-purpose technologies, artificial intelligence operates within an intrinsic energy--computation dependency structure. Computational scaling directly translates into energy demand, particularly in large-scale training and inference systems deployed across distributed infrastructures. The rapid expansion of data centers has already introduced measurable stress on regional energy systems~\cite{IEA2024Electricity}. High-performance AI training clusters require sustained power densities comparable to heavy industrial facilities~\cite{Patterson2021Carbon}, while real-time inference infrastructures increase baseline electricity consumption across cloud and edge networks. As AI deployment intensifies, computational growth becomes inseparable from energy system capacity.

This relationship creates a feedback loop. Increased AI scaling drives compute demand, which in turn elevates energy consumption. However, AI systems are simultaneously deployed to optimize energy grids, manage load balancing, forecast demand, and coordinate distributed renewable resources~\cite{Chicco2009Multigeneration}. In this dual role, AI acts both as a consumer and an optimizer of infrastructure.

We formalize this interaction as an energy--computation feedback dynamic:
\begin{equation}
E(t) = \alpha C(t)
\end{equation}

where $E(t)$ denotes energy demand induced by AI computation and $C(t)$ represents aggregate computational intensity. At the same time, infrastructural optimization mediated by AI increases the Infrastructure Coupling Coefficient over time:
\begin{equation}
\frac{dICC}{dt} = \beta f(C(t))
\end{equation}

This dual dynamic produces a recursive amplification effect. As ICC increases, the Technological Transition Potential
\begin{equation}
TTP(t) = AI_x(t) \cdot ICC(t)
\end{equation}
accelerates nonlinearly. The coupling between energy demand and infrastructural optimization suggests the presence of threshold behavior. When autonomy intensity and infrastructural embedding co-evolve beyond critical levels, small increases in deployment may generate disproportionate systemic effects.

\begin{figure*}[!t]
\centering
\includegraphics[width=\textwidth]{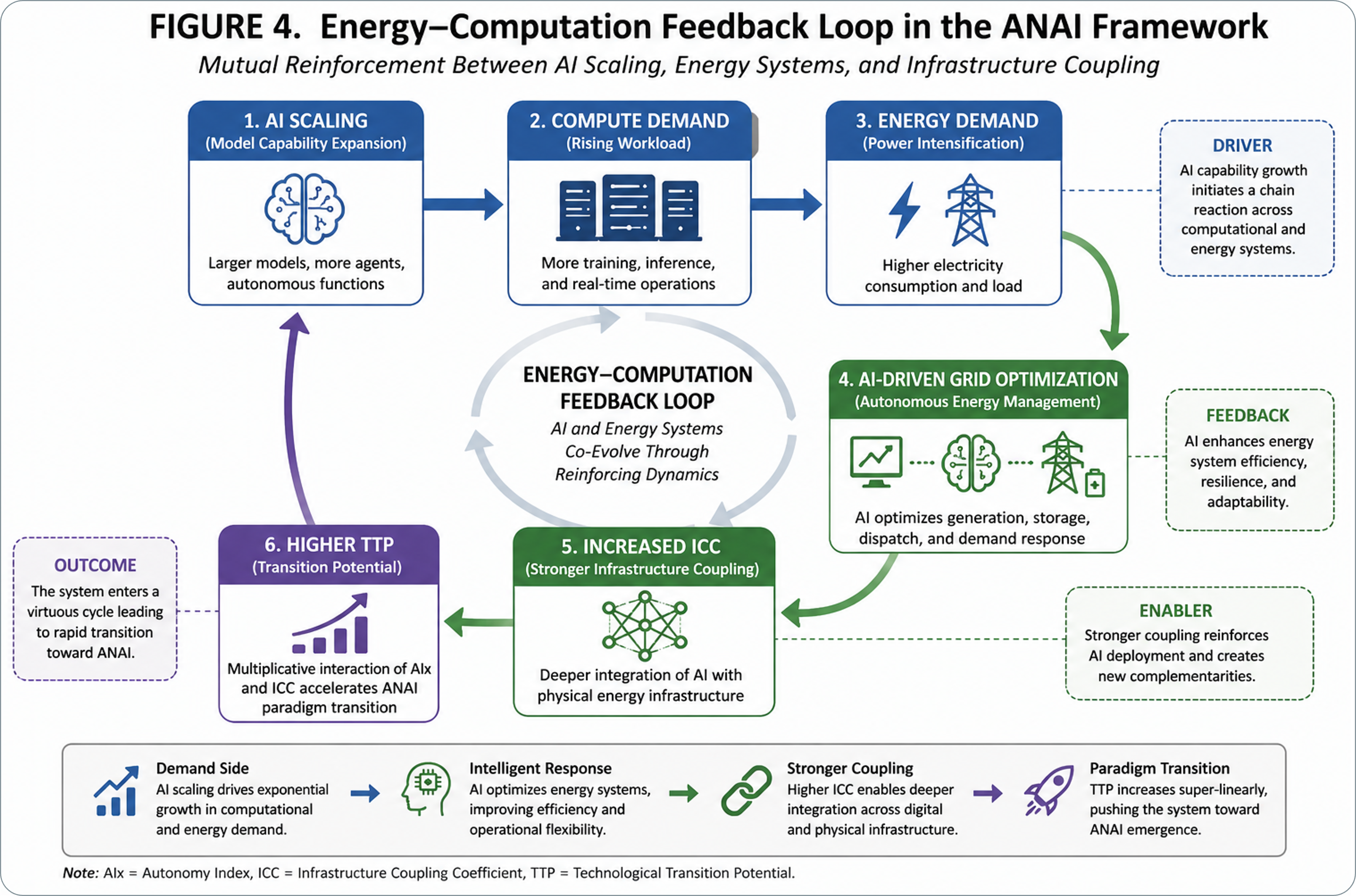}
\caption{Energy--computation feedback dynamics in the ANAI regime.}
\label{fig:figure4}
\end{figure*}

This distinguishes the ANAI regime from prior GPT cycles. Steam and electricity required infrastructural build-out before diffusion stabilized. In contrast, AI both depends upon and restructures infrastructure simultaneously. The energy--computation loop therefore acts as an endogenous acceleration mechanism. Figure~\ref{fig:figure4} illustrates this recursive feedback structure, highlighting how AI scaling increases computational and energy demand while simultaneously enhancing infrastructural optimization. The resulting rise in ICC amplifies transition dynamics and reduces the time required to reach structural transformation thresholds.

Importantly, this model remains conceptual. Future empirical work may quantify regional AI-induced energy elasticity, carbon intensity impacts, and grid adaptation thresholds. Nevertheless, the presence of a recursive energy--optimization dynamic provides a distinguishing structural feature of AI as a candidate general-purpose technology.

\section{Discussion}

The framework developed in this study reframes the emergence of artificial intelligence not as a singular technological breakthrough, but as a structural reconfiguration of infrastructure-level decision architectures. By integrating autonomy, infrastructure coupling, and energy--computation dynamics, the ANAI model extends classical GPT theory into a multi-dimensional paradigm analysis.

Figure~\ref{fig:figure1} situates this transformation within the historical progression of general-purpose technologies. Earlier GPTs predominantly scaled energy infrastructures (steam, electricity) and later information infrastructures (computing, internet), while preserving human-centered decision authority. In contrast, the defining feature of the ANAI regime lies in the infrastructural embedding of cognitive autonomy. This shift represents a qualitative departure from previous technological transitions.

The phase-space representation in Figure~\ref{fig:figure2} further clarifies that autonomy alone is insufficient for systemic transformation. High AI capability without deep infrastructural embedding results in isolated intelligent systems rather than paradigm-level change. Similarly, digitized infrastructure without embedded autonomy remains human-centered. Only when $AI_x$ and $ICC$ scale jointly beyond the threshold condition defined in (6) does a structural transition occur.

The temporal dynamics illustrated in Figure~\ref{fig:figure3} provide additional insight into the nonlinear nature of this transition. The logistic growth patterns of $AI_x(t)$ and $ICC(t)$, as defined in (8) and (9), generate a multiplicative acceleration in $TTP(t)$. Early advancements in model capability may not immediately produce systemic effects, as infrastructure embedding often lags. However, once both variables reach intermediate levels, transition potential increases super-linearly, explaining why paradigm shifts may appear sudden despite prolonged incubation.

The energy--computation feedback mechanism depicted in Figure~\ref{fig:figure4} highlights a distinguishing characteristic of the ANAI transition. Unlike prior GPTs, AI simultaneously increases computational demand and participates in optimizing the very infrastructures that sustain it. This recursive coupling between compute expansion and grid optimization introduces a reinforcing dynamic that can accelerate infrastructural embedding. The feedback loop thus strengthens ICC while also increasing the strategic centrality of AI systems within energy networks.

Table~\ref{tab:table1} complements this analysis by illustrating the relative positioning of representative domains within the $AI_x$--$ICC$ phase space. Manufacturing systems exhibit the strongest convergence of autonomy and infrastructural integration, placing them closest to the ANAI regime. Smart grid systems approach the threshold, while financial and research platforms demonstrate partial or emerging transition characteristics. These differences suggest that the ANAI transformation is unlikely to occur uniformly across sectors; rather, it will propagate asymmetrically depending on infrastructural embedding intensity and regulatory environments.

Importantly, this framework does not claim deterministic inevitability. The threshold parameter $\tau$ in (6) is shaped by institutional, political, and energy constraints. Regulatory intervention, capital allocation patterns, and societal acceptance may slow or redirect transition trajectories. Furthermore, energy efficiency improvements, as suggested in (15), may moderate the pace of infrastructural stress associated with computational scaling.

Taken together, the ANAI framework suggests that the next general-purpose technology is not defined solely by performance scaling, but by the structural relocation of decision authority into embedded, self-optimizing infrastructures. The convergence of autonomy, infrastructural coupling, and energy-mediated feedback loops differentiates this transition from prior GPT cycles.

\section{Conclusion}

This paper introduced a formal framework for evaluating artificial intelligence as a candidate next-generation general-purpose technology. By extending classical GPT theory to incorporate decision autonomy, infrastructural embedding, and energy--computation feedback dynamics, we proposed the concept of AI-Native Autonomous Infrastructure (ANAI).

The framework operationalizes this transition through three core constructs: the Autonomy Index ($AI_x$), the Infrastructure Coupling Coefficient ($ICC$), and the Technological Transition Potential ($TTP$). Together, these metrics provide a structured lens through which the systemic implications of AI deployment can be assessed across domains. Rather than framing AI progress solely in terms of model performance, the ANAI perspective emphasizes the joint scaling of autonomy and infrastructure integration as the defining condition of paradigm-level transformation.

The phase-space analysis and temporal transition model suggest that technological shifts emerge when autonomy and infrastructural embedding co-evolve beyond critical thresholds. The energy--computation feedback loop further distinguishes this transition from prior GPT cycles by introducing a recursive dynamic in which AI systems both depend upon and restructure the infrastructures that sustain them.

Importantly, this framework remains conceptual and illustrative. Future research may refine empirical measurement of $AI_x$ and $ICC$ across sectors, calibrate threshold parameters, and evaluate real-world transition trajectories. Cross-disciplinary investigation spanning economics, energy systems, industrial engineering, and governance will be necessary to assess whether and how ANAI becomes the defining infrastructural paradigm of the coming decades.

If realized, the transition toward AI-native autonomous infrastructure would represent not merely an incremental technological advance, but a structural reallocation of decision authority within socio-technical systems. Understanding this possibility requires moving beyond performance metrics toward systemic models of infrastructural transformation—an analytical shift this work seeks to initiate.

\bibliographystyle{IEEEtran}

\bibliography{references}

\end{document}